\documentclass[a4paper,11pt]{article}
\usepackage{pos}
\usepackage[english]{babel}
\usepackage{mathtools}
\usepackage{pos}
\usepackage{lipsum}
\usepackage{wrapfig}
\usepackage{csquotes}
\usepackage{caption}
\usepackage{subcaption}
\usepackage{siunitx}
\usepackage{titlesec}
\usepackage{hyphenat}
\usepackage{tikz}
\usetikzlibrary{positioning}
\usepackage{cleveref}
\usepackage{subcaption}

\title{Impact of Open Heavy-Flavour Production on nCTEQ Nuclear PDFs}

\author*[a]{Jan Wissmann}
\emailAdd{jan.wissmann@uni-muenster.de}
\author[a]{Tomáš Ježo}
\author[a]{Michael Klasen}
\author[b]{Ingo Schienbein}
\author[c]{Hubert Spiesberger}

\affiliation[a]{Institut für Theoretische Physik, Universität Münster,\\
Wilhelm-Klemm-Straße 9, Münster, Germany}

\affiliation[b]{Laboratoire de Physique Subatomique et de Cosmologie, Université Grenoble-Alpes,\\
CNRS/IN2P3, 53 Avenue des Martyrs, 38026 Grenoble, France}

\affiliation[c]{PRISMA+ Cluster of Excellence,
Institute for Nuclear Physics and Institute of Physics,\\
Johannes Gutenberg-University, Staudingerweg 9, 55099 Mainz, Germany}

\abstract{We present nCTEQ26OHF, a global analysis of nuclear parton distribution functions (nPDFs) within the nCTEQ framework. 
In this analysis, constrasting the main nCTEQ26 release, we replace the Crystal-Ball function, an effective matrix element that is used in nCTEQ26 for open heavy-flavor (OHF) and quarkonium production, by an SACOT GM-VFNS calculation for the OHF predictions. The resulting nCTEQ26OHF PDFs are compatible with the nCTEQ26 PDFs within uncertainties, implying that the OHF data alone provide ample constraint.
}

\FullConference{The 33rd International Workshop on Deep Inelastic Scattering and Related Subjects (DIS2026)\\
4–8 May 2026\\
Bologna, Italy\\}

\hypersetup{
    pdfauthor={Jan Wissmann},
    pdftitle={PineAPPL Grids of Open Heavy-Flavor Production in the GM-VFNS},
    pdfsubject={DIS2024 Proceedings}
}

\NewDocumentCommand{\comment}{m}{\iffalse#1\fi}

\makeatletter
\newcommand*{\centerfloat}{%
  \parindent \z@
  \leftskip \z@ \@plus 1fil \@minus \textwidth
  \rightskip\leftskip
  \parfillskip \z@skip}
\makeatother

\titlespacing{\section}{0pt}{4pt plus 4pt minus 2pt}{4pt plus 2pt minus 2pt}

\renewcommand{\hookAfterAbstract}{%
\begin{tikzpicture}[remember picture,overlay]
    \node [below left=5mm and 4mm of current page.north east] (preprint) {MS-TP-26-26};
\end{tikzpicture}%
}

\begin{document}
\maketitle

\section{Introduction}

Since the discovery of the EMC effect \cite{EuropeanMuon:1983wih}, it is known that nuclei are not comprised of free protons and neutrons, but their structure receives nuclear modifications. Thus, nuclear parton distribution functions (nPDFs) are needed in addition to PDFs of free nucleons to describe, for example, proton-nucleus collisions at high-energy colliders. Because PDFs are non-perturbative objects, but QCD factorization predicts their universality across processes, their dependence on the momentum fraction \(x\) is fitted to data in global QCD analyses such as the recent EPPS21 \cite{Eskola:2021nhw}, nNNPDF3.0 \cite{AbdulKhalek:2022fyi} and nCTEQ15HQ \cite{Duwentaster:2022kpv}. Following the latter, we present nCTEQ26 \cite{ncteq26}, the most recent nCTEQ global analysis that combines our releases since nCTEQ15 \cite{Kovarik:2015cma}.

The recent nPDF analyses show that open heavy-flavor (OHF) production data is of great importance to constrain the gluon PDF at low \(x\). In nCTEQ15HQ and nCTEQ26, we use the Crystal-Ball (CB) function \cite{komPairProductionPsi2011,lansbergAutomatedToolEvaluate2017,kusinaGluonShadowingHeavyFlavor2018} to calculate theory predictions of OHF and quarkonium production. It parametrizes an effective matrix element that is fitted to proton-proton data and then used in proton-lead predictions. To scrutinize this method, we perform an alternative fit that we call nCTEQ26OHF, in which we replace the CB approach by a pQCD SACOT calculation \cite{kniehlInclusive$DPlusminus$2005,Kniehl:2005mk,Kniehl:2015fla}.

\section{nCTEQ26 Setup}

\begin{figure}[b]
    \centering
    \includegraphics[width=0.7\textwidth]{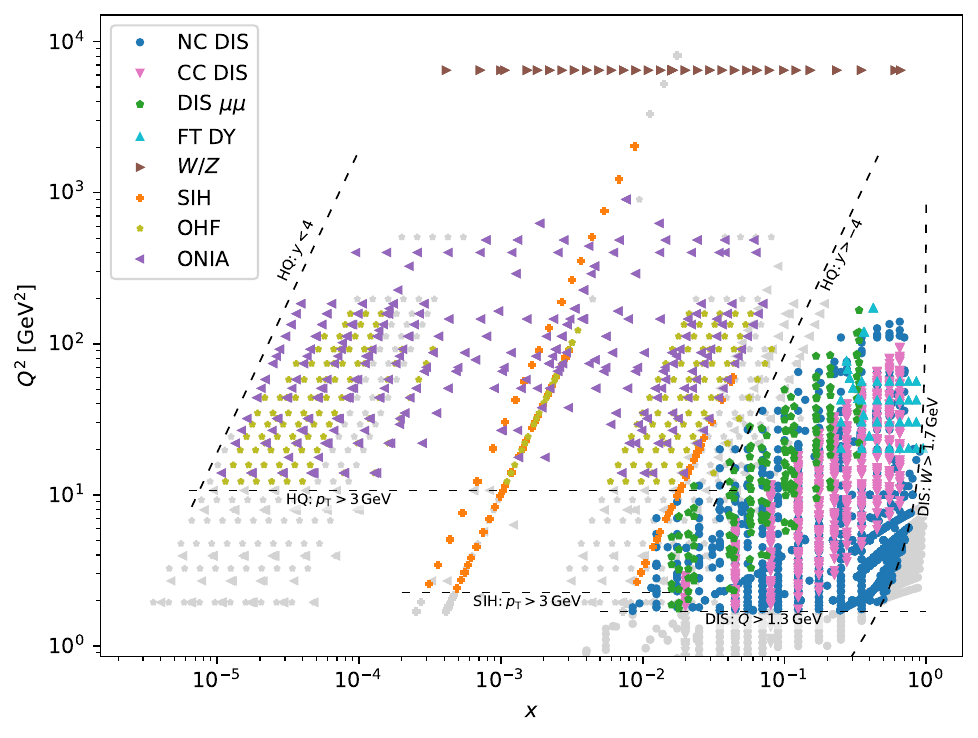}
    \caption{Kinematic coverage of the nCTEQ26 data. In nCTEQ26OHF we leave out the quarkonium data here labeled \texttt{ONIA}.}
    \label{fig:kinematic-coverage}
\end{figure}

First, we will review updates to the nCTEQ26 methodology after nCTEQ15.
Key improvements include, but are not limited to, the addition of JLab neutral-current deep-inelastic scattering (DIS) data and less strict DIS cuts (nCTEQ15HIX \cite{Segarra:2020gtj}), addition of \(W/Z\) and heavy-quark production data (nCTEQ15HQ \cite{Duwentaster:2022kpv}) as well as addition of charged-current DIS data (nCTEQ15\(\nu\) \cite{Muzakka:2022wey}).
This expands the nCTEQ26 data selection to 3598 points compared to 740 points in nCTEQ15, significantly increasing our kinematic coverage in the low-\(x\) region especially due to LHC OHF and quarkonium production data (see \cref{fig:kinematic-coverage}).
As in the previous nCTEQ global analyses, we start from a baseline proton PDF parametrization by adding an \(A\)-dependent part to its parameters.
In contrast to nCTEQ15, we choose CJ15 \cite{accardiConstraintsLargeParton2016} as our new proton baseline in nCTEQ26, thus using \(u_v\), \(d_v\), \(\bar{u} + \bar{d}\), \(\bar{d}/\bar{u}\), \(s + \bar{s}\) and \(g\) as our flavor basis and switching to the parametrization
\begin{equation}
    x f(x, Q_0) = c_0 x^{c_1} (1 - x)^{c_2} (1 + c_3 \sqrt{x} + c_4 x) e^{c_5 \sqrt{x}} \label{eq:parametrization}
\end{equation}
at the initial scale \(Q_0 = \qty{1.3}{\GeV}\). The \(d_v\) PDF receives \(u_v\)-dependent admixture in addition to \cref{eq:parametrization}, and \(\bar{d}/\bar{u}\) is parametrized differently as it is a ratio.
The \(c_5\) parameter is not present in the CJ15 analysis, but we add it to gain more flexibility in the low- to mid-\(x\) region of \(s + \bar{s}\), which is also the only observable for which we open the \(c_5\) parameter.

All parameters receive the \(A\)-dependence
\begin{equation}
    c_j (A) = p_j + a_j \ln A + b_j \ln^2 A \, .
\end{equation}
Of the 68 fittable \(a_j\) and \(b_j\) parameters, we open 35 and set the remaining 33 parameters to zero, which effectively inherits the respective proton parameters. This is a stark increase compared to nCTEQ15, where only 16 of 64 parameters were actually opened, with the wealth of new data making it possible to fit more and more parameters.
We fit 19 nuclei, gaining boron in comparison to nCTEQ15. The analysis is performed at NLO QCD in the SACOT heavy-quark scheme. For the error analysis, we use the usual Hessian method with the tolerance \(T = 50\) and dynamic rescaling of the Hessian eigenvectors in each positive and negative eigenvector direction. For a more complete presentation of the setup, we refer to \cite{ncteq26}.

\section{nCTEQ26OHF Setup}
\begin{figure}
    \vspace{-8pt}
    \centerfloat{\includegraphics[width=0.9\textwidth]{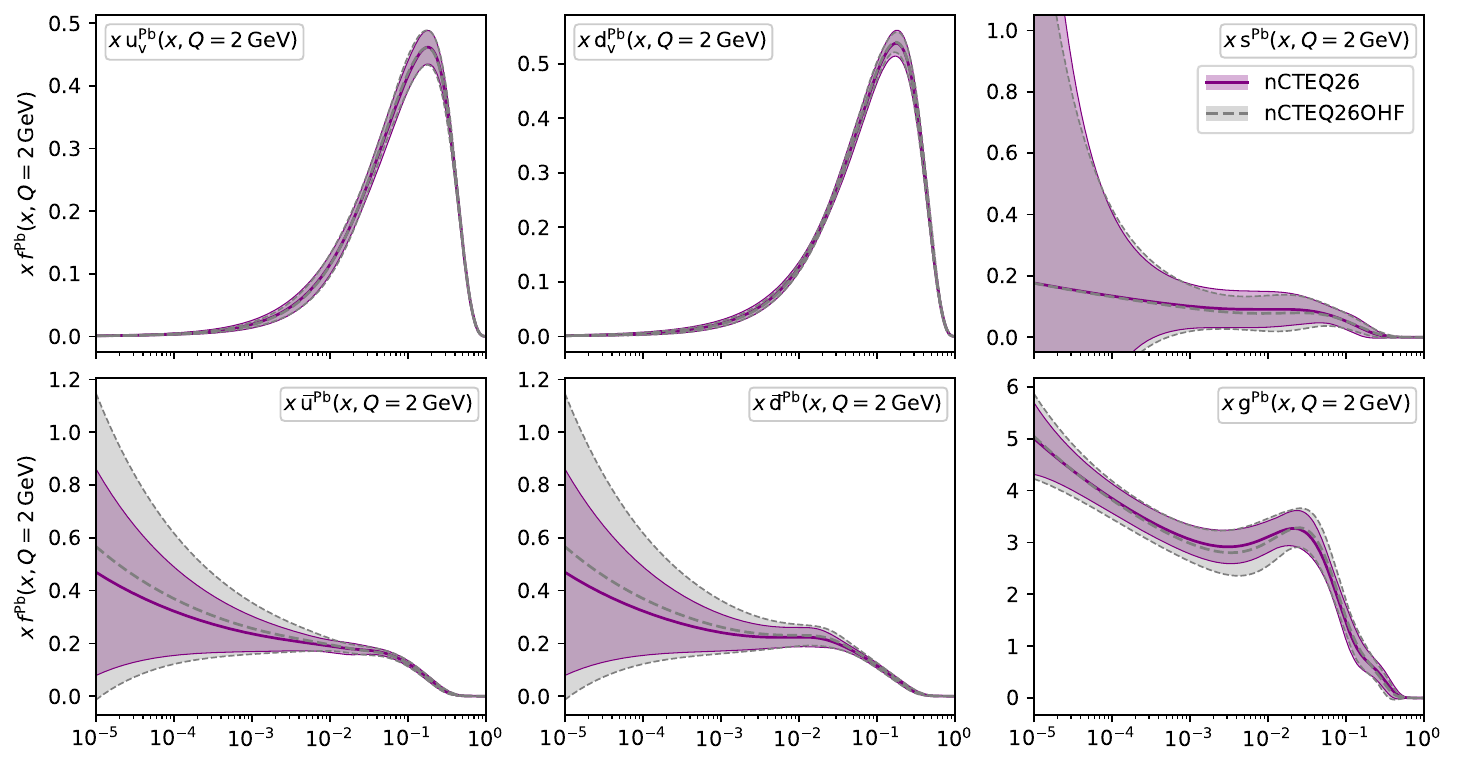}}
    \vspace{-8pt}
    \caption{Comparison of nCTEQ26 and nCTEQ26OHF PDFs.}
    \label{fig:pdfs}
\end{figure}
In the nCTEQ26OHF fit, we assess the compatibility of the CB function with a pQCD calculation in the nCTEQ26 context.
Thus, the nCTEQ26OHF fit retains the setup of nCTEQ26, with the only difference that the quarkonium data is left out and the CB OHF predictions are replaced by an SACOT calculation \cite{kniehlInclusive$DPlusminus$2005,Kniehl:2005mk,Kniehl:2015fla}, using the KKKS08 fragmentation functions \cite{Kneesch:2007ey}, in the form of PineAPPL grids \cite{Carrazza:2020gss,Wissmann:2024bca,Jezo:2026adf}.
In particular, we use the same data as in nCTEQ26 for the other processes and employ the same cuts. Following previous works \cite{Benzke:2017yjn,benzkeMesonProductionGeneralmass2019}, we choose %
a coefficient different from unity in the scale choice \(\mu_i = \xi_i \sqrt{p_\mathrm{T}^2 + 4 m_c^2}\), namely \((\xi_r, \xi_f, \xi_f') = (1, 0.6, 2.5)\) for the renormalization, initial-state factorization and final-state factorization (fragmentation) scales.
In our work, the latter are tuned to the same set of proton-proton baseline data that are used to fit the CB parameters, and then used in the proton-lead fit (to which they are not tuned).
Still, this scale choice gives a much better description of the proton-lead data and compensates for the fact that, currently, no theory uncertainties are used in nCTEQ global analyses.

\section{nCTEQ26OHF Results}
\begin{figure}
    \vspace{-8pt}
    \centerfloat{\includegraphics[width=0.68\textwidth]{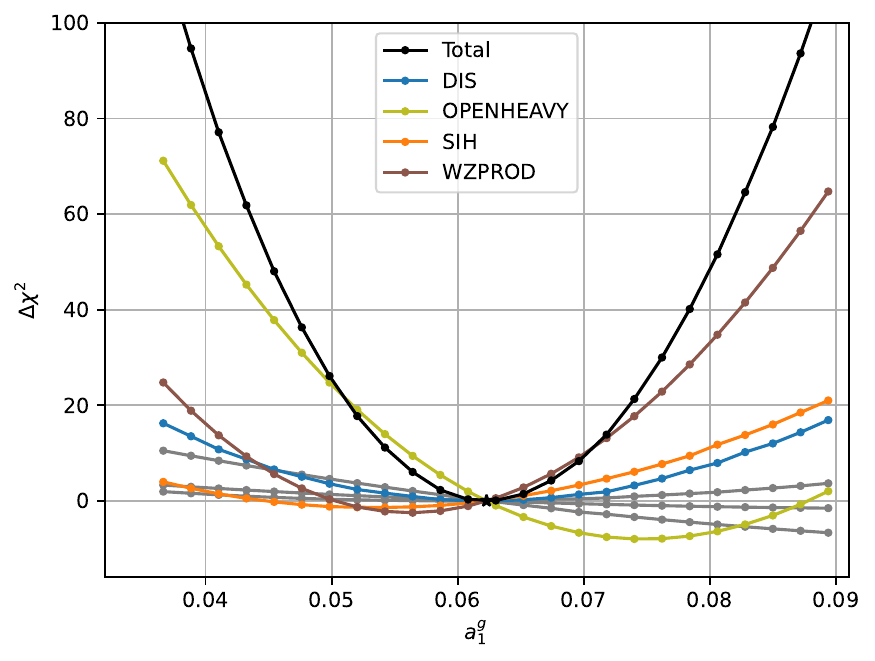}}
    \vspace{-8pt}
    \caption{Scan of the nCTEQ26OHF \(\chi^2\) function along the \(a_1\) gluon parameter, split into the fitted processes. Each curve is shifted along the \(y\) axis such that it intersects the origin at the minimum of the total \(\chi^2\). Highlighted are the processes that show the most constraint.}
    \label{fig:scan}
\end{figure}
In \cref{fig:pdfs}, we compare the nCTEQ26 PDFs and nCTEQ26OHF PDFs.
It is immediately apparent that the centrals agree within uncertainties for all observables.
The gluon uncertainty bands are slightly larger in nCTEQ26OHF than in nCTEQ26, which is expected since removing the quarkonium data weakens the constraint on the gluon.
One could expect the nCTEQ26OHF gluon uncertainty band to be even wider.
But, as was shown for example in the nNNPDF3.0 analyses with and without \(D^0\) production \cite{AbdulKhalek:2022fyi}, the OHF data still has a lot of constraint, especially in the low-\(x\) region.
We can see this also in \cref{fig:scan}, showing the \(\chi^2\) function of the nCTEQ26OHF fit along the \(a_1\) parameter direction, the latter 
\begin{figure}
    \centering
    \begin{subfigure}[t]{0.49\textwidth}
        \centering
        \includegraphics[width=\textwidth]{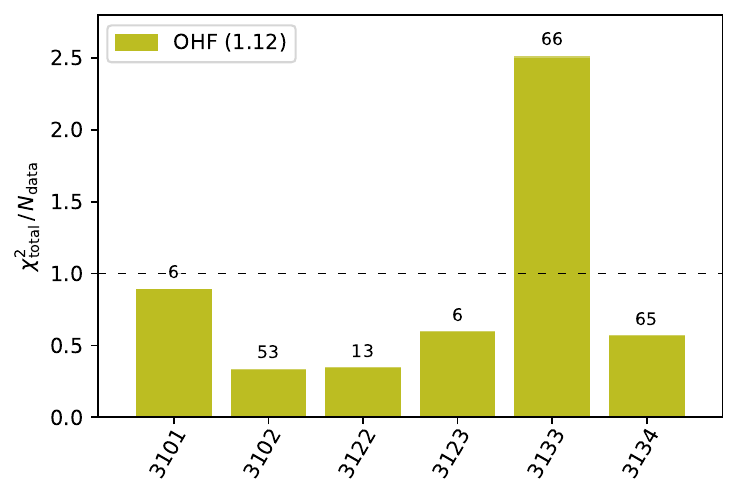}
        \caption{nCTEQ26}
        \label{fig:chi2-ncteq26}
    \end{subfigure}
    \begin{subfigure}[t]{0.49\textwidth}
        \includegraphics[width=\textwidth]{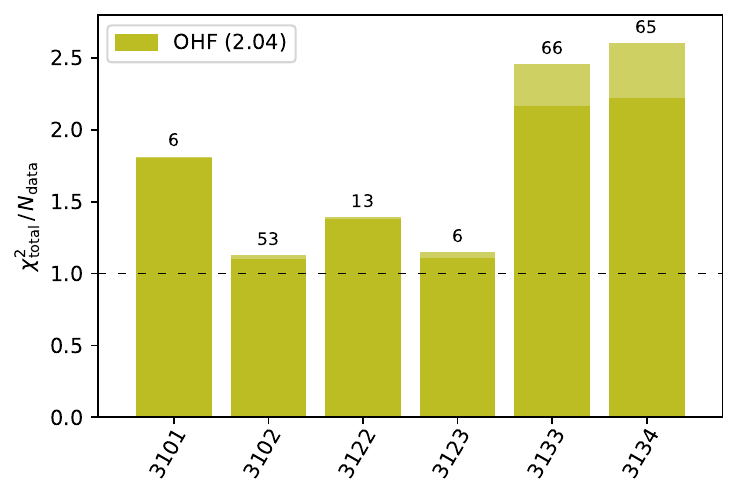}
        \caption{nCTEQ26OHF}
        \label{fig:chi2-ncteq26ohf}
    \end{subfigure}
    \caption{\(\chi^2\) breakdown for the open heavy-flavor data included in nCTEQ26 and nCTEQ26OHF.}
    \label{fig:chi2}
\end{figure}
having the most effect at low \(x\).
The OHF data shows constraint in the negative parameter direction, whereas the \(W/Z\) production data constrains the \(a_1^g\) parameter on the positive side.
This flanking behavior narrows the \(\chi^2\) function, leading to a decrease in gluon uncertainty.
An additional reason is the inclusion of CB fit parameter uncertainties in the nCTEQ26 fit, which increases the nCTEQ26 gluon uncertainties.
Since the CB function is not used in nCTEQ26OHF, these uncertainties are of course absent.
One could include uncertainties on the tuned scale coefficients \(\xi_i\), but we leave this exercise for future work.

In \cref{fig:chi2}, we display the nCTEQ26 and nCTEQ26OHF \(\chi^2\) values of the OHF data.
Except for data set 3133 (LHCb \(D^0\) at \(\sqrt{s} = \qty{8.16}{\GeV}\) \cite{LHCb:2022dmh}, \(y > 0\)), the \(\chi^2\) values are generally larger in the nCTEQ26OHF fit than in the nCTEQ26 fit.
Again, this can be traced back to the CB parameter uncertainties for some datasets, which decrease the \(\chi^2\) in the nCTEQ26 fit even though the description of the data is not necessarily worse.

\section{Conclusion}
In summary, we presented improvements of the new nCTEQ26 global analysis over the nCTEQ15 analysis \cite{Kovarik:2015cma} and select intermediate releases \cite{Segarra:2020gtj,Duwentaster:2022kpv,Muzakka:2022wey}, in total increasing the number of data points from 740 to 3598. Furthermore, we showed the nCTEQ26OHF fit, in which we replaced the CB function by an SACOT calculation in OHF and leaves out the quarkonium data entirely. Despite using less data and having an increased \(\chi^2\), the nCTEQ26OHF PDFs are in good agreement with nCTEQ26. This shows that the OHF data alone hold significant constraint and the CB function is broadly compatible with the SACOT calculation used in our fit.

\section{Acknowledgments}
We thank our nCTEQ colleagues for useful discussions and valuable feedback. Work in Münster was funded by the Deutsche Forschungsgemeinschaft (DFG) through the Research Training Group GRK 2149.

\bibliography{references}

\end{document}